\documentclass[letterpaper]{article}

\usepackage[T1]{fontenc}

\usepackage{geometry}
\usepackage{setspace}

\usepackage{hyperref}
\usepackage[style = chem-acs,  backend=biber]{biblatex}
\usepackage{graphicx}
\usepackage{wrapfig}
\usepackage{float}
\newfloat{scheme}{htbp}{los}
\floatname{scheme}{Scheme}
\floatname{chart}{Chart}
\newfloat{graph}{htbp}{loh}
\usepackage{tocloft}
\usepackage{chemformula} % Formulas using \ch{}
\usepackage[version = 4]{mhchem} % Formulas using \ce{}

\usepackage{authblk}
\author[1]{\normalsize Connor Heimig}
\author[1]{\normalsize Jonas Biechteler}
\author[2]{\normalsize Cristina Cruciano}
\author[2]{\normalsize Armando Genco}
\author[1]{\normalsize Thomas Weber}
\author[1]{\normalsize Michael Hirler}
\author[1]{\normalsize Dmytro Gryb}
\author[1]{\normalsize Alexander A. Antonov}
\author[1,3]{\normalsize Leonardo de S. Menezes}
\author[2,4]{\normalsize Gianluca Valentini}
\author[4]{\normalsize Cristian Manzoni}
\author[2,4]{\normalsize Giulio Cerullo}
\author[5,6]{\normalsize Stefan A. Maier}
\author[1]{\normalsize Luca Sortino$^*$}
\author[1]{\normalsize Andreas Tittl$^\dag$}

\affil[1]{\small Chair in Hybrid Nanosystems, Nanoinstitute Munich, Faculty of Physics, Ludwig-Maximilians-Universit{\"a}t M{\"u}nchen, Munich, Germany}
\affil[2]{\small Dipartimento di Fisica, Politecnico di Milano, 20133 Milano, Italy}
\affil[3]{\small Departamento de Física, Universidade Federal de Pernambuco, 50670-901 Recife-PE, Brazil}
\affil[4]{\small IFN-CNR, Istituto di Fotonica e Nanotecnologie, Milano, Italy}
\affil[5]{\small School of Physics and Astronomy, Monash University, Clayton, Victoria, Australia}
\affil[6]{\small Department of Physics, Imperial College London, London, UK}

\title{\textbf{Integration of 2D Materials in Radial van der Waals Heterostructure Metasurfaces}}
\date{\small Email: $^\ast$~Luca.Sortino@physik.uni-muenchen.de / $^\dag$~Andreas.Tittl@physik.uni-muenchen.de}

\begin{document}

\maketitle

\input{insbox}
  \normalsize
\begin{abstract}
  %\InsertBoxR{10}{\includegraphics[scale=1]{Figures/TOC Figure.png}}
  \normalsize
  \noindent Two-dimensional semiconductors, such as monolayer transition metal dichalcogenides (TMDC), exhibit strong excitonic transitions at room temperature and offer a unique platform for exploring light-matter interactions in nanoscale photonic systems. In this work, we demonstrate a compact and polarization-invariant photonic metasurface, fabricated from hexagonal boron-nitride (hBN) and based on radial bound states in the continuum (BIC), which are formed by radially distributed pairs of structurally asymmetric resonators. The metasurface employs multiple symmetry-breaking perturbations to support high quality-($Q$-)factor resonances within a footprint smaller than 8~$\times$~8~\textmu m$^2$ -- one-sixth of the area of previous approaches. Compared to established hBN metasurface designs, the radial geometry furthermore achieves significantly higher \textit{Q}-factors with a reduced footprint.
By integrating the hBN photonic structure with a WS\textsubscript{2} monolayer, we observe enhanced photoluminescence when its resonance is spectrally aligned with the exciton resonance, accompanied by signatures of discrete momentum-space patterns that identify the orbital-angular-momentum-carrying ring eigenmodes. These features persist over a wide range of excitation powers and show minimal linewidth broadening, indicating robust and spatially modulated exciton-photon coupling. This work establishes a scalable approach for generating hybrid photonic-excitonic states with momentum-space structure, offering new opportunities for exciton localization, valley emission, spatially programmable light-matter interaction in two-dimensional material platforms and compact luminescent devices based on 2D material-integrated metasurfaces.

\end{abstract}

%\subsubsection*{Keywords}
%Van der Waals Materials,
%Metasurfaces,
%Heterostructures, 
%Bound States in the Continuum,
%%low-index photonics
%Exciton-photon coupling

\clearpage
\section*{Introduction}

Among the diverse family of 2D materials, monloayer semiconducting TMDCs such as MoS$_2$, MoSe$_2$, and WS$_2$ stand out for strongly bound excitons with large oscillator strengths, direct band gaps in the visible to near-infrared range~\cite{mak2010atomically, wang2018colloquium, chernikov2014exciton}, and valley-dependent optical selection rules~\cite{xiao2012coupled,ye2014probing, dufferwiel2015exciton}, making them ideal for exploring exciton physics and photonic device concepts~\cite{Novoselov2016, Geim2013, Bonaccorso2010}. 
Coupling TMDC excitons to photonic nanostructures can enhance luminescence~\cite{li2024giant}, control nonlinear processes~\cite{seyler2015electrical}, and even reach strong light-matter coupling regimes~\cite{flatten2016room, lundt2016room}.
To harness these properties in photonic systems, 2D materials are commonly embedded in van der Waals (vdW) heterostructures, where atomically flat interfaces and low-defect dielectric environments improve optical quality. Hexagonal boron nitride (hBN) is central to this platform, providing a chemically inert and wide-bandgap ($\sim$ 6 eV) encapsulation medium that reduces exciton inhomogeneous broadening and ensures stable interfaces~\cite{Dean2010, Wang2013, ajayi2017approaching}. 

Beyond its passive role, hBN has also emerged as an active photonic material, with a low refractive index ($n \sim 2.1$) and optical transparency across the visible and near infrared~\cite{biechteler2025fabrication}. 
%In the mid-infrared (mid-IR), its anisotropic phonon modes can support phonon-polaritons, hybrid excitations of light and lattice vibrations that propagate with deeply subwavelength confinement and low losses~\cite{caldwell2014sub, dai2015subdiffractional}. Complementing material advances, photonic metasurfaces have become a central strategy for manipulating light at the subwavelength. 
Patterned hBN nanostructures have enabled infrared metasurfaces that support Mie resonances, directional emission, and BICs~\cite{giles2018ultralow, li2018boron, kuhner2023high}. These planar arrays of nanoresonators can control phase, amplitude, and polarization of light, and when designed to support BICs, can exhibit high-\textit{Q} resonances despite being embedded in the continuum of radiative modes~\cite{hsu2016bound}. BICs arise from symmetry protection or modal interference and do not couple to the far field. By introducing controlled symmetry breaking, quasi-BICs (qBICs) can be transformed from dark modes to leaky modes that couple to the far field while maintaining high \textit{Q}-factors~\cite{koshelev2018asymmetric}. These qBICs have enabled advances in nonlinear optics~\cite{koshelev2019nonlinear}, biosensing~\cite{tittl2018imaging}, and coupling to quantum emitters and 2D semiconductors~\cite{do2024room, al2021enhanced}.

However, most existing qBIC metasurfaces are designed as periodic lattices with square or rectangular unit cells, which typically results in polarization sensitivity. To address these limitations, radial qBICs have been proposed as a polarization-invariant and compact alternative~\cite{kuhner2022radial}. Their reduced footprint and polarization robustness make them attractive for integration with monolayer TMDCs, for enhanced light-matter coupling, and in applications requiring localized control of the electric field distribution, such as electro- or thermo-optic modulation~\cite{li2017single, gan20222d}.

Rotationally symmetric nanophotonic resonators, ranging from single-resonator discs and microrings to whispering-gallery-mode structures, have been extensively studied due to their ability to confine light with high \textit{Q}-factors and well-defined angular momentum~\cite{little1997microring, vahala2003optical}. More recently, this framework has been extended through lattice-level perturbations and nanoscale unit-cell engineering, giving rise to novel platforms that combine global angular modes with additional photonic resonances, originating from the engineered nanoscale unit cell.
Representative implementations include asymmetric microrings, radial metasurfaces, and circular photonic crystals~\cite{wu2025revealing, kuhner2022radial, ma2024circular}.

In this work, we design and realize a compact hBN-based radial vdW metasurface and demonstrate its integration with a WS\textsubscript{2} monolayer. The geometry is tuned to the WS\textsubscript{2} A-exciton and engineered to support a high-\textit{Q} qBIC together with a discrete ladder of ring-resonator eigenmodes carrying orbital angular momentum (OAM). Full-wave and eigenmode simulations show that the qBIC provides a radiative channel that couples these otherwise guided ring modes to the far field, producing periodically modulated dispersions observed in momentum space. Back-focal-plane (Fourier-space) Photoluminescence (PL) reveals discrete momentum-space patterns consistent with this OAM ladder, with a clear one-to-one correspondence to observed transmittance behavior. 
Overall, the radial architecture delivers polarization-invariant access to high-\textit{Q} modes within a 8-\textmu m$^2$ footprint, providing a practical route to compact and OAM-carrying light-matter interfaces in 2D materials. Potential applications include structured light-exciton emission, near-field beam shaping, and integration with tunable or active vdW photonic elements.

\clearpage

\section*{Results}
\subsection*{Optimization of Radial vdW Metasurfaces from hBN}
\begin{figure}[h!]
    \centering
    \includegraphics[width=1\linewidth]{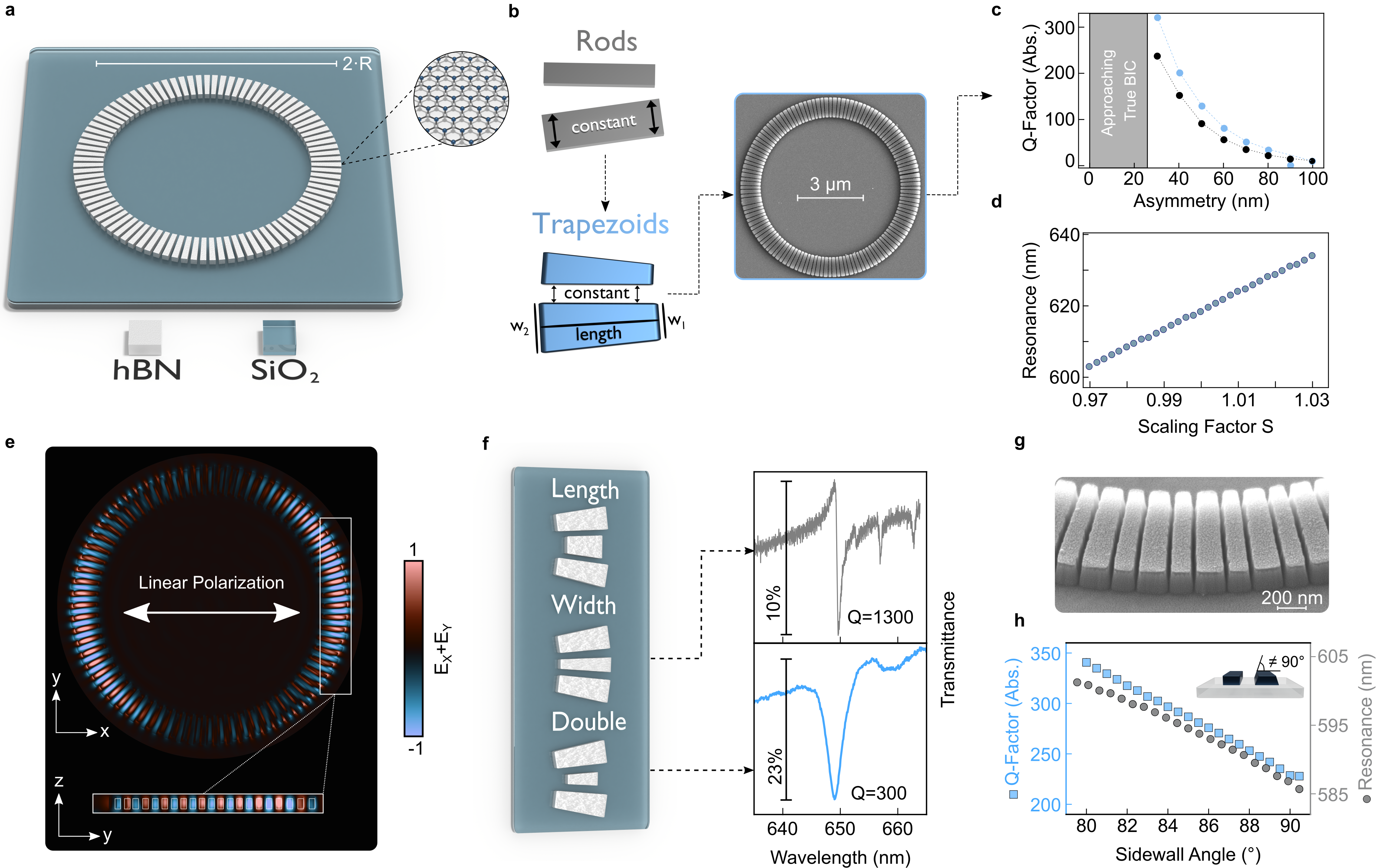}
    \caption{Optimization of radial qBICs in hBN. 
    \textbf{a} Radial qBIC structure with diameter 2$\cdot$\textit{R} fabricated from hBN on a SiO\textsubscript{2} substrate. The inset shows the hBN lattice structure. 
    \textbf{b} When moving from a rod-type to a trapezoid-type unit cell (the SEM image shows a fabricated trapezoidal geometry sample with width asymmetry) the fixed parameter is no longer the width of the individual resonators but the gap between them (see Supplementary Note 2), enabling a 
    \textbf{c} 20\% increase in \textit{Q}-factor in numerical simulations. For this, length asymmetry as introduced in previous work is used~\cite{kuhner2022radial}. 
    \textbf{d} The spectral position of the resonance is tuned via a scaling factor applied to all parameters other than height and number of unit cells. 
    \textbf{e} Sum of real parts of in-plane electric fields for radial qBIC in xy- and yz-plane. 
    \textbf{f} Comparison of different asymmetry approaches with corresponding experimental verification, allowing for both \textit{Q}-factor and signal optimization. 
    \textbf{g} SEM of the fabricated structure revealing slanted sidewalls. 
    \textbf{h} Simulations demonstrating resonance tuning and \textit{Q}-factor control through the inclusion of slanted sidewalls.
    }
    \label{fig:fig1}
\end{figure}

Conceptually, radial metasurfaces are constructed by extracting a single row of resonators from the 2D array which constitutes a planar BIC metasurface and bending it into a circular geometry along the coupling direction, i.e., the short axis of the rods which dominates their optical interaction~\cite{golz2024revealing}. This transforms the 1D resonator chain into a ring-shaped structure while preserving the coupling orientation tangentially along the circumference. 
The radial qBIC structure consists of \textit{N} = 68 double-element unit cells arranged in a circular geometry with radius \textit{R} = 3.95 \textmu m, fabricated from hBN on a SiO\textsubscript{2} substrate (Fig.~ \ref{fig:fig1}a). The resonators are separated by a uniform gap of \textit{d} = 60 nm, with individual bar lengths of \textit{l} = 920 nm. The hBN lattice structure, shown in the inset of Fig.~\ref{fig:fig1}a, provides the fundamental building block for the structure with a height of \textit{h}=165 nm.
Previous implementations of radial qBIC structures relied on simple rectangular (rod-type) resonators with fixed dimensions, directly adapted from such 1D chains. However, while a linear chain maintains constant spacing, bending it into a ring naturally introduces a radially increasing gap between resonators, leading to non-ideal coupling. We introduce a geometric improvement by transitioning from rod-type to trapezoid-type unit cells (Fig.~\ref{fig:fig1}b). This modification restores the constant gap between resonators and increases the resonator volume compared to fixed-width designs, an essential improvement for sustaining strong optical responses in the low-index hBN system, which otherwise operates near the grating mode cutoff.
In the trapezoid configuration, the resonator width is no longer fixed, but varies continuously across the radial position. Each resonator is characterized by inner and outer widths (w\textsubscript{1} and w\textsubscript{2}) that are calculated based on the optimized parameter set (see Supplementary Note 2). 
The trapezoid geometry leads to enhanced \textit{Q}-factors, as evidenced by the approximate 20\% increase observed in numerical simulations when transitioning from rod to trapezoid configurations (Fig.~\ref{fig:fig1}c), both employing the length asymmetry introduced in previous work~\cite{kuhner2022radial}.

To achieve spectral tunability of the resonances, we introduce a unified scaling factor \textit{S} that is applied to all structural parameters simultaneously, except for the resonator height (set by the thickness of the hBN flake) and the number of unit cells. This preserves the relative proportions of the design while enabling controlled tuning across a broad wavelength range (Fig.~\ref{fig:fig1}d and Supplementary Note 3). In contrast, previous work applied scaling only to the radius~\cite{kuhner2022radial}, which indeed shifts the resonance but at the cost of altering the relative geometries.
The in-plane electric field distribution of the radial qBIC under linearly polarized excitation reveals the fundamental coupling mechanism within the structure (Fig.~\ref{fig:fig1}e). The field map shows opposing dipoles formed in adjacent unit cells, which couple throughout the entire ring structure. The field enhancement is strongest in regions parallel or nearly parallel to the incident polarization direction, yet still appearing throughout the full structure. Due to this radial symmetry, rotating either the structure or the polarization direction produces equivalent responses, demonstrating the polarization-independent nature of the radial~qBIC~geometry.
Previous approaches for qBIC generation in radial structures relied solely on length asymmetry to break the symmetry and achieve finite \textit{Q}-factors (of approximately 300) in the visible spectral range.~\cite{kuhner2022radial} However, this approach inherently reduces the effective area of mode confinement, limiting the achievable performance. We introduce two further asymmetry control strategies that address this limitation (Fig.~\ref{fig:fig1}f).
The asymmetries are defined as relative values compared to the unperturbed symmetric system (see Supplementary Note 2 for detailed explanation). First, we implement width asymmetry as an alternative to length modulation. This approach achieves sizable \textit{Q}-factors of \textit{Q} = 1300, as demonstrated in the experimental transmittance spectrum obtained with 30\% width asymmetry (Fig.~\ref{fig:fig1}f).
Secondly, we develop a double asymmetry approach that combines both length and width variations. The double asymmetry spectrum shown combines the 30\% width asymmetry with an additional 20\% length asymmetry. This technique enables higher overall asymmetries without making individual resonators too thin, to maintain structural stability, or too short, to preserve adequate mode confinement area. The doubly asymmetric configuration achieves a higher transmission modulation at the expense of a reduced \textit{Q}-factor ($Q$ = 300, Fig.~\ref{fig:fig1}f). 
The reduced Q-factor reflects stronger radiative losses induced by the higher asymmetry. While higher-Q modes store energy longer, they are more sensitive to parasitic losses, such as material absorption and fabrication imperfections, which can diminish the measurable modulation depth. 
This flexibility allows the system to be tailored depending on whether maximum \textit{Q}-factor or optimal signal modulation is prioritized for specific applications. The heterostructure discussed in later sections employs this double-asymmetry. Beyond the in-plane geometric optimizations, we introduce an additional out-of-plane tuning mechanism through controlled sidewall etching angles (Fig.~\ref{fig:fig1}g and h). This approach addresses fundamental fabrication limitations by circumventing minimum achievable gap sizes through slanted sidewall profiles. The anisotropic etching not only provides an alternative pathway to reduce effective gap dimensions but also serves as a powerful tool for both \textit{Q}-factor enhancement and spectral tuning.
The relationship between sidewall angle and optical performance is demonstrated through  simulated characterization, showing substantial \textit{Q}-factor improvements concurrent with resonance wavelength shifts. Throughout this work, a sidewall angle of 85$^{\circ}$ is employed, representing an optimal compromise between fabrication reliability and optical performance. 

\subsection*{Radial qBIC-Mediated Coupling of Ring Resonator Modes to the Far Field}

\begin{figure}[h!]
    \centering
    \includegraphics[width=1\linewidth]{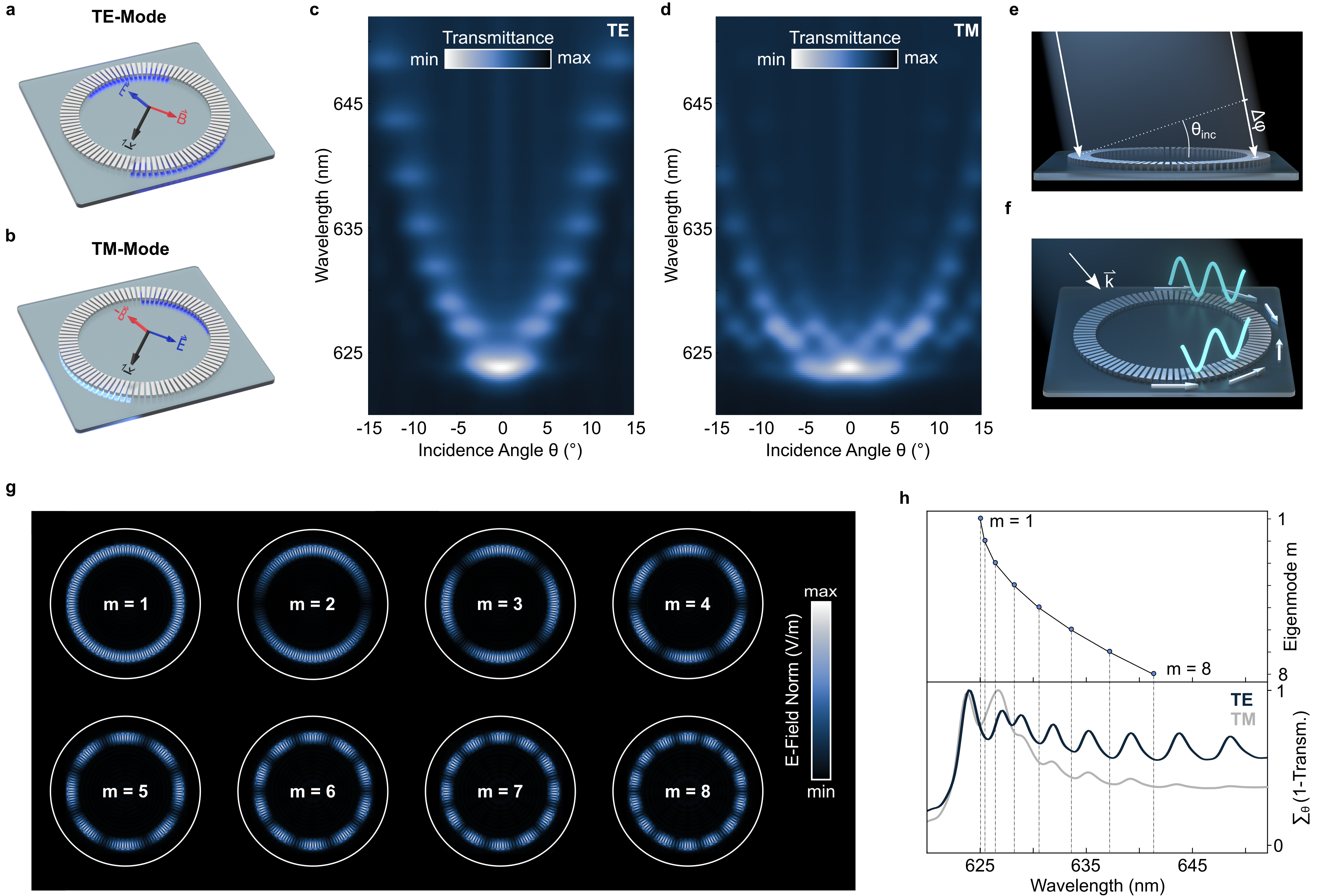}
    \caption{Radial qBIC in k-space. 
     \textbf{a (b)} Sketch of TE (TM) mode constituent fields in respect to radial qBIC structure for oblique incidence. A phase difference is accumulated in the magnetic (electric) component. 
     \textbf{c} Simulated transmittance dependency for oblique incidence in TE-mode. The resonance trend exhibits a periodically modulated parabolic shape. 
     \textbf{d} Simulated transmittance dependency for oblique incidence in TM-mode. In this case, the resonance trend exhibits a shape constituted by a series of periodically modulated parabolas. 
     \textbf{e} Sketch of the resulting phase difference $\Delta\phi$ at opposing sides of the radial qBIC structure. 
     \textbf{f} Sketch of angled light, launching propagating waves at the respective ring poles with opposing propagation direction, resulting in a standing wave. 
     \textbf{g} E-Field Norm for the first eight eigenmodes (mode number 1 to 8) of the radial qBIC structure. 
     \textbf{h} Spectral position of the respective eigenmodes correlated with (1 - T) summed up over all angles of incidence for both TE and TM. 
    }
    \label{fig:fig2}
\end{figure}

The angular response of radial qBICs reveals a rich interplay between localized unit-cell resonances and collective ring eigenmodes. To systematically study this behavior, we define the incident polarization states relative to the angled excitation geometry (Fig.~\ref{fig:fig2}a and b). For oblique incidence, TE-mode excitation corresponds to the electric field oriented perpendicular to the plane of incidence, while TM-mode maintains the electric field within the incidence plane. This distinction becomes crucial when analyzing the angle-dependent transmittance spectrum.
Under TE-mode excitation, the angular transmittance dispersion exhibits a periodically modulated parabolic profile (Fig.~\ref{fig:fig2}c). While the underlying parabolic envelope is consistent with previous demonstrations of qBIC behavior~\cite{sortino2025atomic, jiang2023general}, the varying modulation represents a new feature arising from the specific geometry and excitation conditions of our radial qBIC system.
This behavior becomes even more pronounced under TM-mode excitation (Fig.~\ref{fig:fig2}d), where the dispersion reveals not only the periodically modulated parabolic envelope but multiple distinct parabolic branches. This  multi-branch structure indicates the presence of several modes that contribute to the overall optical response, suggesting a rich landscape of different electromagnetic eigenmodes in the system.

The origin of the multiple resonance branches can be understood through a phase accumulation mechanism that occurs during oblique incidence (Fig.~\ref{fig:fig2}e). When the excitation angle deviates from normal incidence, phase differences accumulate across the structure. In TM-mode, this phase accumulation occurs in the electric field component that directly drives the resonant response through electric dipole coupling. Consequently, different phase accumulation conditions create multiple solutions that manifest as distinct resonance branches in the angular dispersion.
The periodically modulated features arise from ring resonator eigenmodes excited under tilted illumination, as illustrated schematically in Fig.~\ref{fig:fig2}f: angled excitation launches counter-propagating waves at opposite ring poles, whose interference produces the standing-wave patterns that define the eigenmodes around the ring.
In conventional microring platforms, such whispering-gallery or OAM eigenmodes are guided resonances, confined by total internal reflection and typically accessed only via evanescent in-plane coupling schemes such as bus waveguides or tapered fibers~\cite{hammer2022resonant}. The large in-plane wavevector of such circulating modes typically cannot couple directly to free-space plane waves, making them effectively dark to far-field excitation and detection.
In our radial qBIC system, however, the leaky qBIC resonance provides a built-in radiative channel. The qBIC breaks symmetry just enough to scatter guided fields into the continuum while retaining a high quality factor, thereby mediating coupling between the normally bound OAM states and free-space radiation. As a result, modes that would remain hidden in standard microrings become directly observable in angle-resolved transmittance and PL. This mechanism effectively transforms the radial qBIC into a far-field interface for the OAM ladder of ring eigenmodes, enabling their study without the need for waveguide coupling or near-field probes.

To substantiate our interpretation of the periodically modulated envelope as ring eigenmodes made radiatively accessible by the qBIC, we compute the first eight eigenmodes of the radial structure using COMSOL (Fig.~\ref{fig:fig2}g). Each solution is characterized by a well-defined OAM $\ell = m$, in line with prior reports on photonic crystal microrings~\cite{wu2025revealing}. The $m = 2$ solution corresponds to the fundamental qBIC mode, while higher-order solutions ($m = 3,4,\ldots,8$) represent distinct OAM states that are normally confined by total internal reflection but become radiatively coupled through the qBIC. Their characteristic $m$-fold azimuthal field patterns directly reflect their angular momentum content.

The correspondence between these eigenmodes and the transmittance behavior of the same structure is demonstrated in Fig.~\ref{fig:fig2}h: the eigenfrequencies calculated for the ring resonator align closely with the oscillatory features in the angle-resolved calculated (1-transmittance) spectra of the same structure. This spectral agreement confirms that the features observed in the angular dispersion indeed arise from a ladder of OAM-carrying eigenmodes. Rather than enhancing only a single resonance, the radial qBIC opens a radiative access channel to these guided modes, which would otherwise remain confined to the near field unless additional outcoupling structures, such as gratings, were introduced.

\subsection*{Radial qBIC vdW-Heterostructure}
\begin{figure}[h]
    \centering
    \includegraphics[width=1\linewidth]{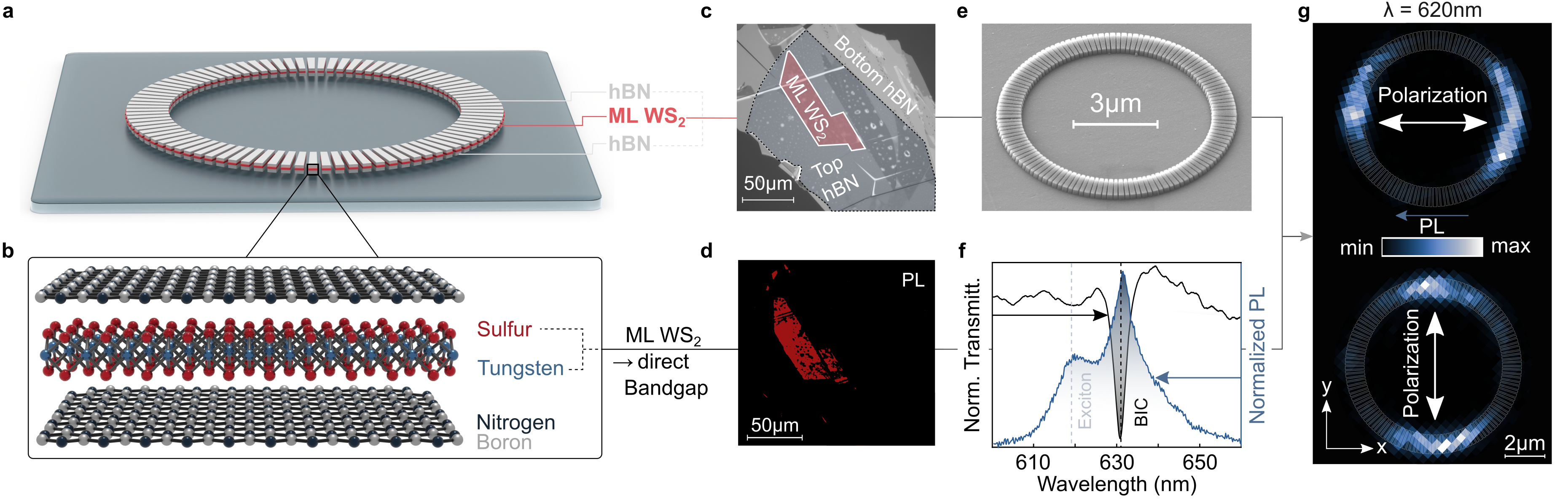}
    \caption{Radial qBIC vdW-heterostructure. 
     \textbf{a} Radial qBIC vdW-heterostructure with a mono-atomic layer (ML) of WS\textsubscript{2} encapsulated by two bulk layers of hBN. 
     \textbf{b} Respective lattice structures of heterostructure components. 
     \textbf{c} Optical microscope image of final heterostructure. Both the bottom and top layer of hBN are 80 nm thick. 
     \textbf{d} PL microscope image of the heterostructure shows signal from the area of the monolayer.  
     \textbf{e} SEM image of the fabricated structure.
     \textbf{f} Normalized Transmittance and PL spectra for radial qBIC with scaling factor $S$ = 1.01. Both the excitonic response and a BIC-driven resonant enhancement are visible in the PL spectrum. 
     \textbf{g} Experimental PL maps of radial qBIC vdW-heterostructure close to the excitonic wavelength ($\lambda$ = 620 nm) show BIC-driven enhancement of PL emission.
    }
    \label{fig:fig3}
\end{figure}

To explore exciton-photon interactions, we fabricated a vdW heterostructure consisting of a WS\textsubscript{2} monolayer sandwiched between two 80~nm thick hBN flakes, and subsequently patterned it into the radial qBIC platform (Fig.~\ref{fig:fig3}a - e). The WS\textsubscript{2} monolayer has a direct bandgap and strong excitonic response at room temperature, resulting in bright room-temperature PL emission (Fig.~\ref{fig:fig3}d). The encapsulation in hBN furthermore provides an atomically flat environment that preserves optical quality and minimizes inhomogeneous broadening.~\cite{sortino2025atomic}
To facilitate the interaction between the radial qBIC and the WS\textsubscript{2} exciton, we fine-tuned the structure's geometry to place the qBIC resonance near 620~nm, spectrally aligning it with the exciton energy. This alignment establishes the conditions under which exciton-photon interaction may be enhanced, potentially entering the strong coupling regime. To probe the polarization symmetry of the qBIC-enhanced emission, we image the PL intensity at the exciton wavelength ($\lambda = 620$~nm) under fixed linear polarization and repeat the measurement after rotating the sample by 90$^\circ$. Because the radial qBIC is isotropic, this is equivalent to rotating the polarization relative to the structure. For clarity, this is depicted as a polarization rotation in Fig.~\ref{fig:fig3}g. The resulting PL maps remain nearly unchanged, in line with the polarization-invariant response of the radial qBIC. 
The PL intensity is enhanced in regions of the structure where the long axis of the resonators is parallel to the excitation, corresponding to the areas with the strongest near fields (Fig.~\ref{fig:fig1}g), which suggests local field concentration and resonant enhancement of excitonic emission.
Spectral measurements further indicate enhanced light-matter interaction. The normalized PL and transmittance spectra for a structure scaled to $S = 1.01$, show a clear spectral overlap between the PL peak and the radial qBIC resonance (Fig.~\ref{fig:fig3}f). This alignment indicates that the qBIC may enhance exciton emission by increasing the local density of optical states and facilitating efficient radiative coupling.  Under the right conditions radial qBIC can generate self-hybridized exciton-polaritons, opening the door for future experimental studies of strong coupling in such radially symmetric systems (see Supplementary Note 5).

\subsection*{k-Space Hyperspectral Imaging of Radial qBIC vdW-Heterostructure}

\begin{figure}[h!]
    \centering
    \includegraphics[width=1\linewidth]{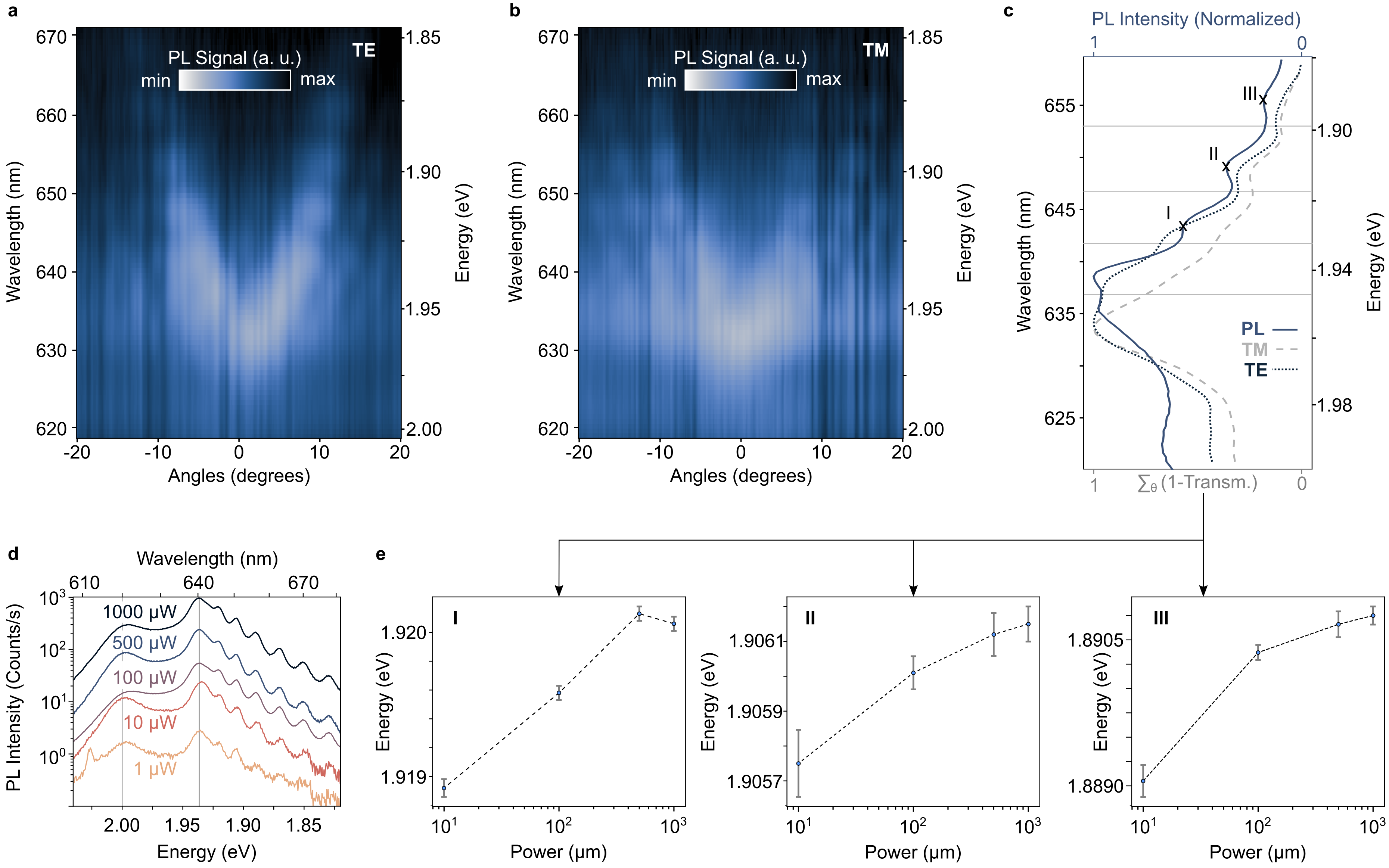}
    \caption{k-Space hyperspectral imaging of radial qBIC vdW-heterostructure. 
    Vertical cross-section of the experimental 3D hyperspectral image showing a periodically modulated parabolic dispersion for \textbf{a} TE and \textbf{b} TM. 
     \textbf{c} PL spectrum correlated with 1-T summed up over all angles of incidence for both TE and TM polarizations. \textbf{d} Power-dependent PL of radial qBIC vdW-heterostructure. 
     \textbf{e} Power-dependence of the peak-position and linewidth of three separate local maxima in the PL spectrum. 
    }
    \label{fig:fig4}
\end{figure}

Figures~\ref{fig:fig4}a and \ref{fig:fig4}b display hyperspectral $k$-space PL images for TE and TM polarizations, respectively. In both polarizations, multiple dispersive features are visible, forming parabolic ripple patterns consistent with eigenmode simulations (Fig.~\ref{fig:fig2}g). For TE (Fig.~\ref{fig:fig4}a), the emission is dominated by a single parabolic branch, whereas in TM (Fig.~\ref{fig:fig4}b) several higher-order contributions appear. To connect this emission to the linear optical response, we integrate the PL intensity over all emission angles, resulting in an angle-integrated PL spectrum. This is directly compared to the normal-incidence (1 – $T$) of the same structure (Fig.~\ref{fig:fig4}c). A clear one-to-one correspondence emerges between the resonance peaks in the PL and the transmittance dips.
Due to electromagnetic reciprocity in linear, time-invariant systems, swapping excitation and collection channels yields the same coupling: emission into angle $\theta$ under normal excitation is reciprocal to illumination from $\theta$ detected at normal incidence. Consequently, our angle-resolved PL (Fig.~\ref{fig:fig4}) mirrors the angle-resolved transmittance (Fig.~\ref{fig:fig2}), and the angle-integrated PL reproduces the transmission dips.
To evaluate the robustness of this behavior under varying excitation conditions, we measure the PL spectrum as a function of pump power (Fig.~\ref{fig:fig4}d). Even at the lowest excitation fluences, the ripple structure remains clearly discernible, indicating that the underlying mode coupling is governed primarily by the geometry and remains stable in the linear regime. As the excitation power increases by three orders of magnitude, these ripple features persist across the spectrum, demonstrating the resilience of the hybrid modes under higher carrier densities and elevated local fields. Fitting the peak positions of selected modes (labeled I, II, and III in Fig.~\ref{fig:fig4}c), we observe a subtle but systematic blueshift in emission energy with increasing pump power (Fig.~4e). This shift is modest, on the order of a few meV, but reproducible, and may be attributed to weak band-filling effects or changes in exciton binding energy due to photoinduced screening in the monolayer.~\cite{chernikov2015population}
Another possible explanation could be light-induced renormalization of excitonic transitions, a known phenomenon in monolayer TMDCs under elevated carrier densities.~\cite{steinhoff2017exciton} Importantly, the linewidths of the hybrid peaks show no dramatic broadening with increasing power, indicating that the system does not enter a loss-dominated or saturated regime within the explored power range. 
Overall, these results are consistent with momentum-resolved coupling between excitonic emission from the monolayer and the structured photonic modes supported by the radial qBIC platform, with both spectral and angular features showing stability across a broad range of excitation conditions. 
Notably, the ability to recover the characteristic patterns associated with OAM modes in the PL response suggests that these angular features persist in the presence of exciton-photon interaction. This indicates that the underlying OAM structure of the qBIC modes may be partially imprinted onto the excitonic emission.

\section*{Discussion}
We employ a periodic, rotationally symmetric architecture with a nanoscale perturbation motif, enabling discrete angular eigenmodes, symmetry-protected qBICs, and enhanced light-matter interactions. While conceptually derived from microring resonators, our implementation adopts a fully metasurface-based design that combines global symmetry with local control to enable compact and tunable optical functionality.

Compared to earlier radial qBIC demonstrations~\cite{kuhner2022radial}, our structures exhibit markedly improved performance: for the same \textit{Q}-factor, we achieve over 2.5$\times$ greater spectral modulation, and for fixed modulation depth, more than a fourfold increase in \textit{Q}-factor, even in a low-index system based on hBN (previous work used silicon). Although this comes with a modest increase in device footprint relative to high-index implementations, our structures remain significantly more compact than typical metasurface approaches. These initial hBN-based qBIC designs required patterned areas exceeding $20 \times 20$~\textmu m$^2$ ~\cite{kuhner2023high}, whereas our radial qBICs occupy a circular footprint of only $\sim$4~\textmu m radius -- less than one-sixth of the area - while leaving the central region unpatterned. This void can be leveraged for spectral multiplexing, emitter integration, or additional photonic functionality. Importantly, these advantages are not accompanied by a drop in performance; our optimized radial geometry reaches \textit{Q}-factors up to 1300, nearly four times higher than the $\sim$300 reported for the initial periodic resonator lattice hBN designs~\cite{kuhner2023high}, underscoring the efficiency of this approach in shaping and confining optical modes.
Our results also demonstrate that integrating monolayer WS\textsubscript{2} into the radial qBIC architecture enables clear signatures of light-matter interaction between the excitonic and photonic components. While direct experimental evidence of strong coupling has not yet been achieved, the spectral overlap between PL and the qBIC resonance, together with momentum-resolved PL, suggests enhanced coupling and structured emission. The persistence of ripple-like features across excitation powers, including weak but reproducible power-dependent shifts, supports the view that the photonic mode structure plays an active role in shaping the excitonic response.
Angular eigenmode hybridization has previously been explored in asymmetric microring photonic crystals~\cite{wu2025revealing}, where discrete angular features were confirmed via near- and far-field intensity measurements. However, those studies focused on purely photonic modes outside the visible range and did not investigate excitonic interaction. 
The reappearance of ring eigenmodes, known to carry OAM, in the PL indicates that OAM content is at least partially transferred to, or preserved within, the qBIC-exciton-coupled emission. This points to the possibility of imprinting angular momentum structure onto light generated from 2D materials via nanophotonic symmetry engineering.
Such momentum-structured emission could provide a foundation for more complex spatial and spectral control. In particular, the ring-like, intensity-modulated PL distributions may enable exciton guidance, localization, or spatially varying light-matter coupling within a single device. These effects open potential pathways for exciton trapping or valley-selective routing, where spin, valley, and momentum degrees of freedom could be coherently manipulated. Additionally, the PL far-field response carrying OAM may enable on-chip information encoding.

\section*{Methods}
\subsection*{Numerical simulations}
Simulations of the transmittance spectra for the radial-BIC hBN metasurfaces were conducted using the finite-difference time-domain (FDTD) package of a commercial software (Lumerical Ansys). The refractive index of the SiO$_2$ substrate was set to 1.45, while that of hBN was taken from literature.~\cite{kuhner2022radial}  The full structure was simulated using perfectly matched layer (PML) boundary conditions.
The eigenmodes of the ring were calculated with the eigenfrequency solver of the wave optics module in COMSOL Multiphysics. Here, the refractive index of hBN was assumed to be a constant value of $n = 2.12$, and the entire ring was considered embedded in air. Due to symmetry, only one half of the ring was simulated, with perfect magnetic conductor (PMC) boundary conditions applied at the symmetry plane.

\subsection*{Sample fabrication}
Fused silica substrates were initially cleaned by sonication in acetone at 55$^{\circ}$C, followed by isopropanol to remove any residual acetone. Subsequently, the substrates were treated with O$_2$ plasma to eliminate organic residue and enhance flake adhesion. To facilitate precise global alignment of the flake position on the substrate during subsequent processing, a marker system was created on the substrates using optical lithography (SÜSS Maskaligner MA6). The hBN flakes were mechanically exfoliated from bulk crystals (HQ Graphene) onto the cleaned silica marker substrates. The deposition process was conducted at a temperature of 105 $^{\circ}$C to evaporate moisture and stretch the exfoliation tape, ensuring flattened transferred flakes. The height of the flakes was measured using a profilometer (Bruker Dektak XT) with a stylus having a radius of 2\,\textmu m. 
The radial hBN metasurfaces were fabricated using an electron-beam lithography (EBL) process, followed by lift-off and reactive-ion etching (RIE). The EBL step was carried out using an eLINE Plus (Raith Nanofabrication) with 20\,kV acceleration voltage and a 10\,\textmu m aperture. A single layer of positive-tone AR-P 6200.13 (Allresist) was used as the EBL resist, spin-coated at 1200\,RPM and baked at 175$^{\circ}$ C for five minutes. Espacer 300Z (Showa Denko K.K.) was then subsequently spin-coated onto the sample. The patterned films were developed in amyl acetate, followed by a mixture of methyl isobutyl ketone and isopropyl alcohol (1:9 ratio).
A hardmask consisting of 2\,nm titanium (Ti) and 35\,nm of chromium (Cr) was evaporated onto the sample using electron-beam evaporation and subsequently lifted off overnight in Microposit Remover 1165 (Microresist). This served as an etching mask for the subsequent reactive-ion etching process, using sulfur hexafluoride (SF$_6$) and argon (Ar) gases under 6.0\,mTorr pressure with 300\,W HF and 150\,W ICP power. The Cr part of the hardmask was removed via reactive-ion etching based on chlorine (Cl$_2$) and oxygen (O$_2$) at a pressure of 12\,mTorr with 20\,W HF power and 500\,W ICP power. The Ti part of the hardmask was subsequently removed using a solution of potassium monoiodide and iodine (Sigma-Aldrich). The full fabrication process is depicted in Supplementary Note~6.

\subsection*{Linear Optical Measurements}
The linear transmittance spectra were characterized using a commercially available confocal optical transmittance microscope (Witec alpha 300 series). The samples were illuminated from the bottom using collimated and linearly polarized white light from a broadband halogen lamp (Thorlabs OSL2). The light was subsequently confocally collected with a $50\times$ objective (NA = 0.8) and coupled into a multimode fiber. The collected signal was guided into a spectrometer with a grating groove density of 600\,grooves/mm, where it was dispersed onto a Si-CCD sensor. 

\subsection*{k-space Hyperspectral Microscope}
To perform the k-space hyperspectral PL measurements, we employed a custom hyperspectral microscopy setup (a sketch of this setup is depicted in Supplementary Note~7). The excitation beam is produced by a 532\,nm continuous-wave (CW) laser, coupled to a multimode fiber (core diameter: 100\,\textmu m), whose tip is imaged onto the sample using a collimation lens (not shown). The laser beam is reflected by a dichroic mirror (RazorEdge 532\,nm) towards a $100\times$ objective (NA = 0.75), which is also used for signal collection. 
The PL emitted by the sample passes through the dichroic mirror and a long-pass filter (LP550), used to further suppress residual excitation light. A Fourier lens in the detection path enables imaging of the back focal plane of the objective onto the camera. Before detection, the Translating-Wedge-Based Identical Pulses eNcoding System (TWINS) interferometer is inserted in the optical path.~\cite{brida2012phase}. By varying the delay between the two replicas of the Fourier-space image generated by the TWINS interferometer, we acquire an interferogram for each pixel. The Fourier transform of each interferogram yields the corresponding PL spectrum. 
The final dataset is a three-dimensional datacube (hypercube) of PL intensity as a function of $\theta_x$, $\theta_y$, and energy. Vertical cross-sections of this hypercube reveal the photonic mode dispersion of the sample for different polarizations (see main text), enabled by the built-in polarizers of the TWINS interferometer. These are set at 45$^\circ$ to select the TE (TM) mode along the diagonal (anti-diagonal), respectively.~\cite{genco2022k}

\section*{Associated Content}

\subsubsection*{Supporting Information Available:}
Si to hBN transition, definitions trapezoidal unit cell and relative asymmetry, impact of geometric parameters, scaling factor, numerical investigation of strong coupling, illustration fabrication workflow, sketch experimental setup (PDF)
\section*{Author Information}
\subsubsection*{Corresponding Authors}

\begin{list}{}{\setlength{\leftmargin}{2em}} 
\item \textbf{Luca Sortino} - Chair in Hybrid Nanosystems, Nanoinstitute Munich, Faculty of Physics,
Ludwig-Maximilians-Universität, Munich, Germany. Email: Luca.Sortino@physik.uni-muenchen.de

\item \textbf{Andreas Tittl} - Chair in Hybrid Nanosystems, Nanoinstitute Munich, Faculty of Physics,
Ludwig-Maximilians-Universität, Munich, Germany. Email: Andreas.Tittl@physik.uni-muenchen.de
\end{list}

\subsubsection*{Authors}
\begin{list}{}{\setlength{\leftmargin}{2em}} 
\item \textbf{Connor Heimig} - Chair in Hybrid Nanosystems, Nanoinstitute Munich, Faculty of Physics, Ludwig-Maximilians-Universität, Munich, Germany.  \href{https://orcid.org/0009-0001-6820-2157}{https://orcid.org/0009-0001-6820-2157}
\item \textbf{Jonas Biechteler} - Chair in Hybrid Nanosystems, Nanoinstitute Munich, Faculty of Physics, Ludwig-Maximilians-Universität, Munich, Germany.
\item \textbf{Cristina Cruciano} - Dipartimento di Fisica, Politecnico di Milano, 20133 Milano, Italy.
\item \textbf{Armando Genco} - Dipartimento di Fisica, Politecnico di Milano, 20133 Milano, Italy.
\item \textbf{Thomas Weber} - Chair in Hybrid Nanosystems, Nanoinstitute Munich, Faculty of Physics, Ludwig-Maximilians-Universität, Munich, Germany.
\item \textbf{Michael Hirler} - Chair in Hybrid Nanosystems, Nanoinstitute Munich, Faculty of Physics, Ludwig-Maximilians-Universität, Munich, Germany.
\item \textbf{Dmytro Gryb} - Chair in Hybrid Nanosystems, Nanoinstitute Munich, Faculty of Physics, Ludwig-Maximilians-Universität, Munich, Germany.
\item \textbf{Alexander A. Antonov} - Chair in Hybrid Nanosystems, Nanoinstitute Munich, Faculty of Physics, Ludwig-Maximilians-Universität, Munich, Germany.
\item \textbf{Leonardo de S. Menezes} - Chair in Hybrid Nanosystems, Nanoinstitute Munich, Faculty of Physics, Ludwig-Maximilians-Universität, Munich, Germany.\\
Departamento de Física, Universidade Federal de Pernambuco, 50670-901 Recife-PE, Brazil.
\item \textbf{Gianluca Valentini} - Dipartimento di Fisica, Politecnico di Milano, 20133 Milano, Italy.\\
IFN-CNR, Istituto di Fotonica e Nanotecnologie, Milano, Italy.
\item \textbf{Cristian Manzoni} - IFN-CNR, Istituto di Fotonica e Nanotecnologie, Milano, Italy.
\item \textbf{Giulio Cerullo} - Dipartimento di Fisica, Politecnico di Milano, 20133 Milano, Italy.\\
IFN-CNR, Istituto di Fotonica e Nanotecnologie, Milano, Italy.
\item \textbf{Stefan A. Maier} - School of Physics and Astronomy, Monash University, Clayton, Victoria, Australia. \\
Department of Physics, Imperial College London, London, UK

\end{list}

\subsubsection*{Author contributions}
C.H., J.B., A.T and L.S. conceived the idea and planned the research. C.H. and J.B. contributed to the sample fabrication. C.H., C.C., A.G, L.S.M., G.V., C.M., C.G. and L.S. performed optical measurements. C.H.,  T.W. and M.H. conducted the numerical simulations and data processing. C.H., T.W., D.G., A.A.A. and L.S.M. developed the theoretical background. G.C., S.A.M., L.S. and A.T. supervised the project. All authors contributed to the data analysis and to the writing of the paper.

\subsubsection*{Conflict of interest}
There are no conflicts to declare.

\subsubsection*{Data availability} 
All data needed to evaluate the conclusions in the paper are present in the paper and/or the Supplementary Materials.

\section*{Acknowledgements}
Funded by the European Union (EIC, OMICSENS, 101129734, ERC, METANEXT, 101078018, QUONDENSATE, 101130384). Views and opinions expressed are however those of the author(s) only and do not necessarily reflect those of the European Union or the European Research Council Executive Agency. Neither the European Union nor the granting authority can be held responsible for them. This project was also funded by the Deutsche Forschungsgemeinschaft (DFG, German Research Foundation) under grant numbers EXC 2089/1-390776260 (Germany’s Excellence Strategy) and TI 1063/1 (Emmy Noether Program), the Bavarian program Solar Energies Go Hybrid (SolTech) and the Center for NanoScience (CeNS). S.A.M. additionally acknowledges the Lee-Lucas Chair in Physics. C.C., A.G. and G.C. acknowledge the European Union’s NextGenerationEU Programme with the I-PHOQS Infrastructure [IR0000016, ID D2B8D520, CUP B53C22001750006] “Integrated Infrastructure Initiative in Photonic and Quantum Sciences”.

\printbibliography
\newcommand{\TitleFont}{\fontsize{18}{20}\selectfont\sffamily\bfseries}
\clearpage
\phantomsection
%\pdfbookmark[0]{Supplementary Materials}{supp}  % starts a new top-level node

  \centering

  % --- SI title (prefix + main title) ---
  {\LARGE \textbf{Supporting Information for Integration of 2D Materials in Radial van der Waals Heterostructure Metasurfaces}\par}

  \vspace{1.2cm}

  % --- Authors (with affiliation numbers + equal contrib + corresponding) ---
  {\normalsize
  Connor Heimig\textsuperscript{1},
  Jonas Biechteler\textsuperscript{1},
  Cristina Cruciano\textsuperscript{2},
  Armando Genco\textsuperscript{2},
  Thomas Weber\textsuperscript{1},
  Michael Hirler\textsuperscript{1},
  Dmytro Gryb\textsuperscript{1},
  Alexander A. Antonov\textsuperscript{1},
  Leonardo de S. Menezes \textsuperscript{1,3},
  Gianluca Valentini\textsuperscript{2,4},
  Cristian Manzoni\textsuperscript{4},
  Giulio Cerullo\textsuperscript{2,4},
  Stefan A. Maier\textsuperscript{5,6},
  Luca Sortino\textsuperscript{*,1},
  Andreas Tittl\textsuperscript{\textdagger, 1}\par}

  \vspace{0.8cm}

  % --- Affiliations (numbered) ---
  {\small
  \textsuperscript{1}\,Chair in Hybrid Nanosystems, Nanoinstitute Munich, Faculty of Physics, Ludwig-Maximilians-Universit{\"a}t M{\"u}nchen, Munich, Germany\\
\textsuperscript{2}\,Dipartimento di Fisica, Politecnico di Milano, 20133 Milano, Italy\\
\textsuperscript{3}\,Departamento de Física, Universidade Federal de Pernambuco, 50670-901 Recife-PE, Brazil\\
\textsuperscript{4}\,IFN-CNR, Istituto di Fotonica e Nanotecnologie, Milano, Italy\\
\textsuperscript{5}\,School of Physics and Astronomy, Monash University, Clayton, Victoria, Australia\\
\textsuperscript{6}\,Department of Physics, Imperial College London, London, UK\par}

  \vspace{0.8cm}
    \centering
  {\date{Email: $^\ast$~Luca.Sortino@physik.uni-muenchen.de / $^\dag$~Andreas.Tittl@physik.uni-muenchen.de}
  \par}

  \vspace{1.0cm}

  % --- SI front matter ON THE SAME PAGE ---
    \raggedright
    \normalsize
    \tableofcontents

  \vfill

% Reset figure, table, and equation counters and add 'S' prefix
\setcounter{figure}{0}
\renewcommand{\thefigure}{S\arabic{figure}}

\setcounter{table}{0}
\renewcommand{\thetable}{S\arabic{table}}

\setcounter{equation}{0}
\renewcommand{\theequation}{S\arabic{equation}}

\clearpage
\section{Supplementary Note 1 - Moving from Silicon to hBN}
\begin{figure}[hb]
    \centering
    \includegraphics[width=1\linewidth]{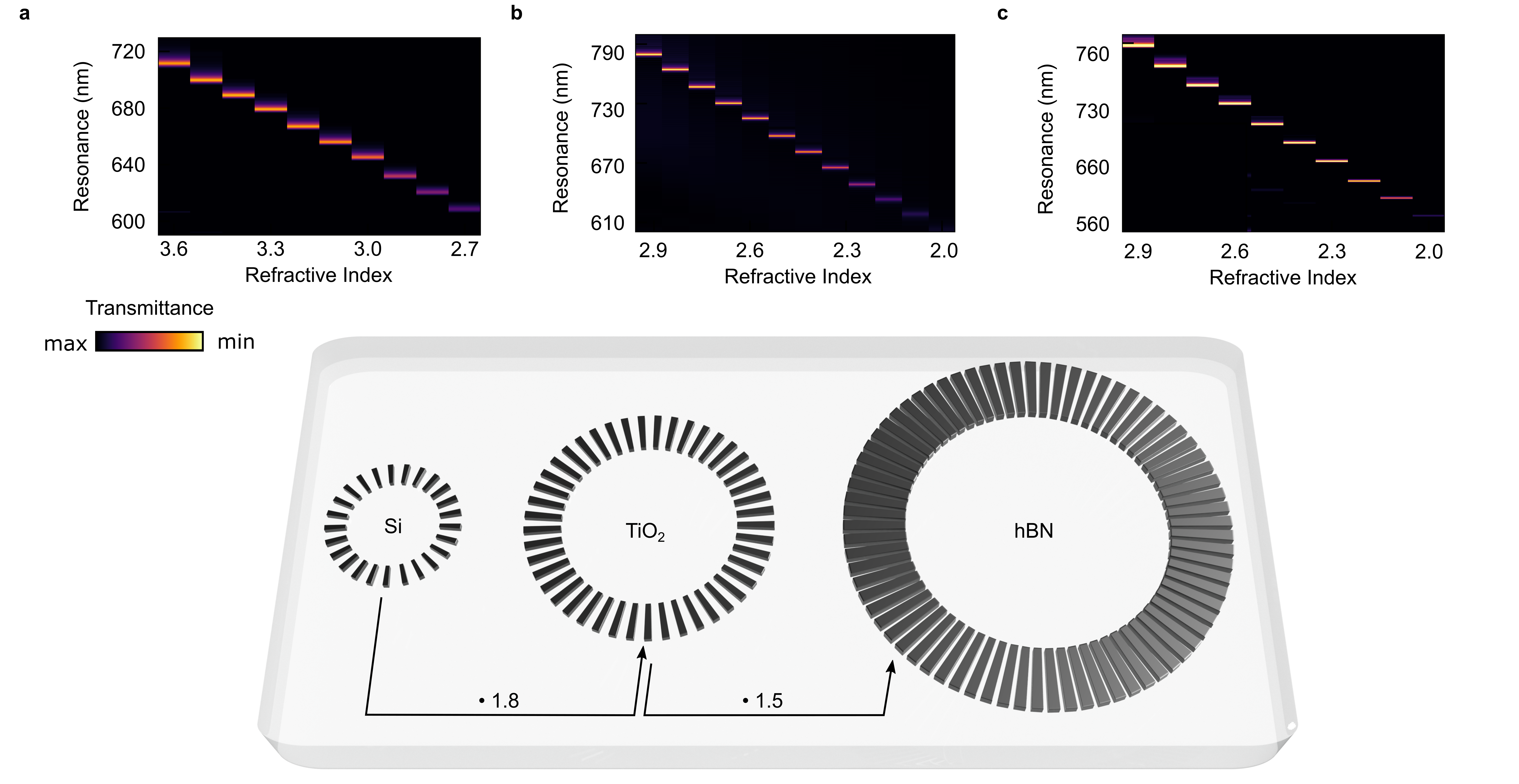}
    \caption{Moving from high refractive index dielectric to low refractive-index. 
    Simulated transmittance for differently scaled radial qBIC structures, starting from the silicon design proposed in literature.~\cite{kuhner2022radial}
    \textbf{a}~The transition from a resonator material with a high refractive index ($n = 3.45$, silicon) to a low-index material such as hBN ($n_{x,y} = 2.1,  n_{z} = 1.6$ in the investigated spectral range) also entails a reduction in refractive-index contrast between the resonators and the substrate (fused silica, $n = 1.45$). 
    \textbf{b}~Therefore, to attain sufficiently strong signal, the dimensions of the ring must first be increased by a factor of approximately 1.8 to recover a qBIC signal around 700\,nm for a refractive index similar to that of TiO$_2$ ($n = 2.6$). 
    \textbf{c}~To achieve a comparable signal for hBN, the dimensions of the structure must be increased again by a factor of 1.5.}
    \label{fig:SitohBN}
\end{figure}

The radial qBIC structure based on silicon features a radius of approximately 1.5\,\textmu m and exhibits a resonance slightly above 700\,nm.~\cite{kuhner2022radial} In a first step, we gradually decrease the refractive index of the resonators, which leads to a blue-shift of the resonance and a reduction in modulation depth (Fig.~\ref{fig:SitohBN}a). Both effects arise from the reduced refractive-index contrast between the resonator and the substrate and are mitigated by uniformly upscaling the entire ring geometry. Increasing the ring radius red-shifts the resonance, while enlarging the resonator dimensions and the number of unit cells increases the total resonator volume, partially compensating for the lower index contrast.
A design scaled by a factor of approximately 1.8 compared to the original silicon structure yields sufficiently strong resonances near 700\,nm at a refractive index of $n = 2.6$, comparable to that of TiO$_2$. Further upscaling by a factor of 1.5, along with additional optimization steps (such as unit cell geometry, see Fig.~1), results in qBIC resonances with improved modulation depth and higher $Q$-factors than the initial silicon-based design.

\clearpage
\section{Supplementary Note 2 - Trapezoidal Unit Cell with Relative Asymmetry}
The exact calculation for the parallel side lengths of the trapezoids ($W_{\mathrm{inner}}$ and $W_{\mathrm{outer}}$, $w_1$ and $w_2$ in the main text) in a symmetric radial qBIC ring uses the following parameters:

\begin{itemize}
    \item $N$: number of unit cells (i.e., a pair of resonators). Each unit cell consists of two trapezoidal resonators, giving a total of $2N$ resonators in the ring.
    \item $R$: ring radius, defined as the distance from the ring center to the center of each trapezoid.
    \item $l$: trapezoid length, i.e., the distance between the two parallel sides.
    \item $d$: desired gap between neighboring resonators.
\end{itemize}

To simplify the calculation, the radial qBIC structure is approximated as a regular $N$-gon (polygon with $N$ sides), instead of a continuous circular arc. The side lengths of the trapezoids are then given by:

\begin{align}
W_{\mathrm{inner}} &= \frac{2\pi \left(R - \frac{l}{2}\right) - 2Nd}{2N}, \\
W_{\mathrm{outer}} &= \frac{2\pi \left(R + \frac{l}{2}\right) - 2Nd}{2N}.
\end{align}

Here, the term in parentheses represents the inner or outer circumference of the polygon, from which the total gap length is subtracted, and then normalized by dividing by the number of resonators.

\vspace{1em}
\noindent \textbf{Relative Asymmetry}\\
\noindent To allow comparison between different radial qBIC designs, a relative asymmetry approach is introduced. Unlike absolute asymmetries, the relative asymmetry expresses differences in trapezoid dimensions as a fraction of the original symmetric dimensions, making designs with different overall scales directly comparable.

Let:
\begin{itemize}
    \item $dL_{\mathrm{rel}}$: relative length asymmetry
    \item $dW_{\mathrm{rel}}$: relative width asymmetry
\end{itemize}

The absolute asymmetries are calculated from the symmetric design as:

\begin{align}
dL &= dL_{\mathrm{rel}} \cdot l, \\
dW_{\mathrm{inner}} &= dW_{\mathrm{rel}} \cdot W_{\mathrm{inner}}, \\
dW_{\mathrm{outer}} &= dW_{\mathrm{rel}} \cdot W_{\mathrm{outer}}.
\end{align}

To implement the asymmetries while maintaining constant gaps, the trapezoid widths are modified. First, the inner and outer widths of the longer resonators (those unaffected by the length asymmetry) are:

\begin{align}
W_{\mathrm{inner,long}} &= \frac{2\pi \left(R - \frac{l}{2}\right) - 2Nd}{2N} + \frac{dW_{\mathrm{inner}}}{2}, \\
W_{\mathrm{outer,long}} &= \frac{2\pi \left(R + \frac{l}{2}\right) - 2Nd}{2N} + \frac{dW_{\mathrm{outer}}}{2}.
\end{align}

Then, for the shorter resonators (those affected by the length asymmetry), the corresponding widths are:

\begin{align}
W_{\mathrm{inner,short}} &= \frac{2\pi \left(R - \frac{l}{2}\right) - 2Nd}{2N} - \frac{dW_{\mathrm{inner}}}{2}, \\
W_{\mathrm{outer,short}} &= \frac{2\pi \left(R + \frac{l}{2}\right) - 2Nd}{2N} - \frac{dW_{\mathrm{outer}}}{2}.
\end{align}

\section{Supplementary Note 3 - Impact of Different Geometric Parameters}
\subsubsection*{Radius}
This parameter allows for spectral tuning of the resonance (Fig.~\ref{fig:GeoParm}a) and can be conceptually compared to the length of a 1D chain. Increasing the radius leads to a red-shift of the resonance, while decreasing it results in a blue-shift. This trend is further influenced by the fact that, in the trapezoidal ring design, an increase in radius also increases the total resonator volume, since the resonator widths are continuously recalculated to maintain a constant total gap volume.

A key advantage of this design, compared to a rod-based approach, is that varying the radius over a reasonable range does not significantly impact the $Q$-factor of the resonance, thus broadening the design flexibility. This robustness is due to the constant gap width preserved by the trapezoidal geometry at any radius, in contrast to rod geometries where the gap width and especially the gap widening are radius-dependent. An explicit discussion of the gap parameter follows below.

Changes in modulation strength with varying radius can be attributed to the changing interaction cross-section between the incident light and the resonator ring.

\subsubsection*{Resonator Height}
This parameter is predetermined by the thickness of the selected hBN flake. The modulation depth is strongly influenced by the resonator height, due to an increased volume available for optical field confinement within a high-index contrast environment. This increased confinement helps mitigate losses from substrate leakage, which explains the substantially reduced signal and $Q$-factor for thinner resonators (Fig.~\ref{fig:GeoParm}b).

\subsubsection*{Resonator Length}
The resonator length refers to the center length of each trapezoid in the symmetric configuration. Increasing this length leads to a red-shift in the resonance position (Fig.~\ref{fig:GeoParm}c), primarily due to the corresponding increase in resonator volume. Additionally, a longer resonator increases the light–matter interaction cross-section, which manifests as stronger modulation in the transmittance spectrum.

\subsubsection*{Gap Width}
The gap width is the fundamental parameter from which the resonator widths are calculated. A smaller gap results in stronger modulation and higher $Q$-factors, as it pushes the structure further into the sub-wavelength regime (Fig.~\ref{fig:GeoParm}d). The associated red-shift observed for narrower gaps is once again a result of increased resonator volume.

Since the resonator width is defined based on the gap width, the limit case of zero-width gaps leads to vanishing resonator width—at which point no mode confinement is possible, and the qBIC far-field response collapses.

\subsubsection*{Number of Unit Cells}
\begin{figure}[ht!]
    \centering
    \includegraphics[scale=0.95]{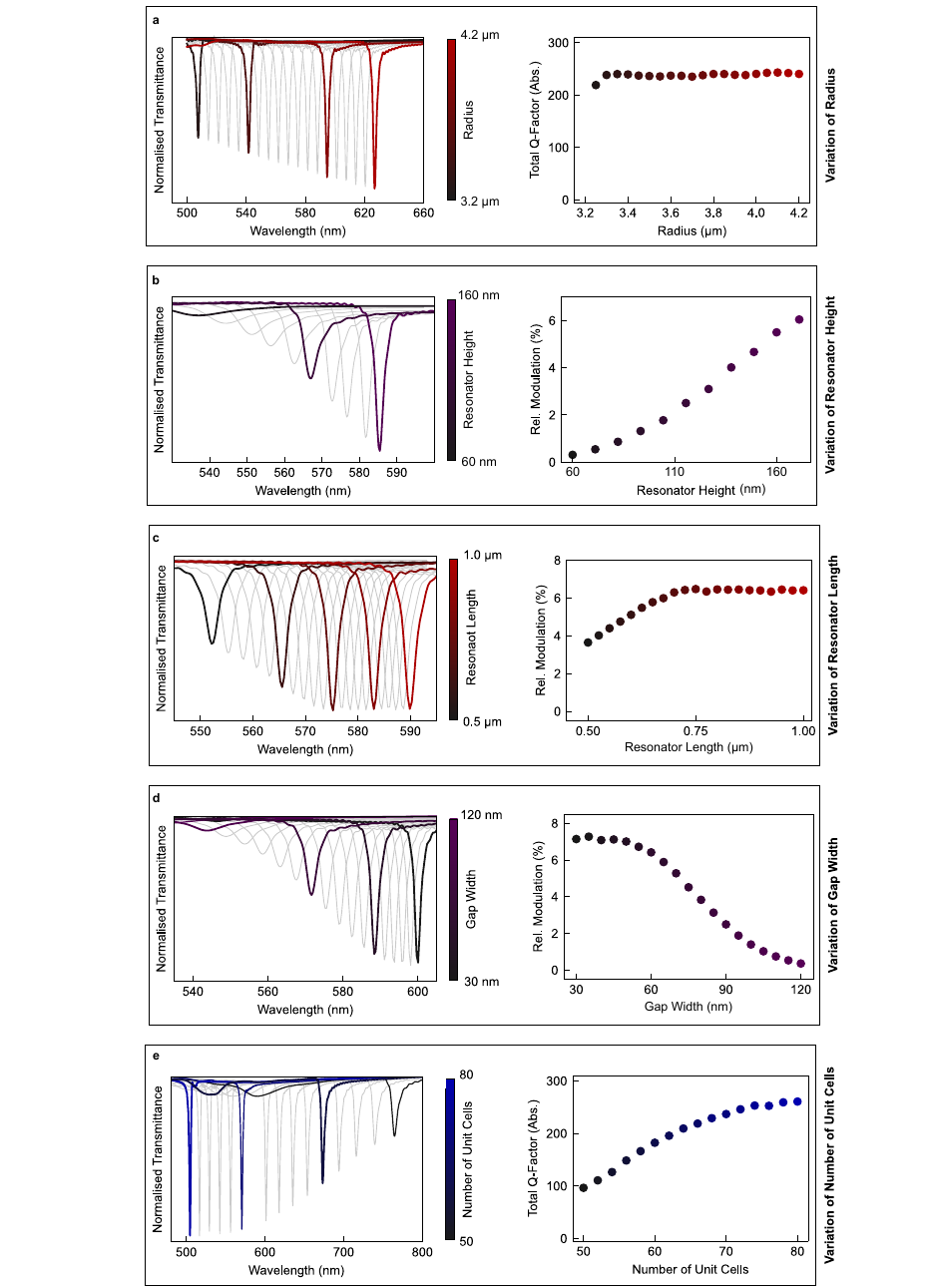}
    \caption{Impact of geometric parameters. 
    Normalized simulated transmittance spectra for perturbation of individual geometric parameters while maintaining all others. The changing parameters are \textbf{a} Radius \textit{R}, \textbf{b} Resonator Height \textit{h}, \textbf{c} Resonator Length \textit{l}, \textbf{d} Gap width \textit{d}, \textbf{e} Number of unit cells \textit{N}.}
    \label{fig:GeoParm}
\end{figure}

Increasing the number of unit cells increases the total number of resonators and, consequently, the number of gaps. Because the gap width is kept constant, increasing the number of resonators changes the ratio between resonator volume and gap volume (i.e., air), leading to a blue-shift in the resonance position. The modulation depth and $Q$-factor increase asymptotically with more unit cells, due to the higher density of potential light confinement sites (Fig.~\ref{fig:GeoParm}e).

However, this improvement eventually saturates. Beyond a certain point, further increasing the number of resonators causes their individual widths to approach zero, again preventing mode confinement and suppressing qBIC formation.

The main practical limitation on the number of resonators is mechanical stability. In fabrication, it has proven advantageous to avoid structures with dimensions below 70\,nm to prevent collapse. This constraint places an upper bound on the number of unit cells in hBN rings resonant around 620\,nm at approximately 68.

\clearpage
\section{Supplementary Note 4 - Introduction of Scaling Factor}
In order to spectrally shift a radial qBIC over a small range, very minor steps in radius would be necessary ($\sim$10~nm). Alternatively, the resonator length or the gap spacing can be changed on the scale of single nanometers. However, if the goal is to recreate a base radial qBIC's spectral response at a different spectral position, tuning only a singular parameter is a not fully optimized approach. Instead, the preferred approach in this work is to multiply every respective parameter of a radial qBIC by a scaling factor.
Hence, all the fabricationally determinable parameters (radius, length, gap, and asymmetry) are multiplied by a scaling factor before running through the outlined calculation of the trapezoid widths. The only excluded parameters are the resonator height, which is assumed to be constant and predetermined by the exfoliated flake's height, and the unit cell number. This is due to the necessity of this always being an even and whole number; therefore, it is kept constant.

\begin{figure}[ht]
    \centering
    \includegraphics[width=1\linewidth]{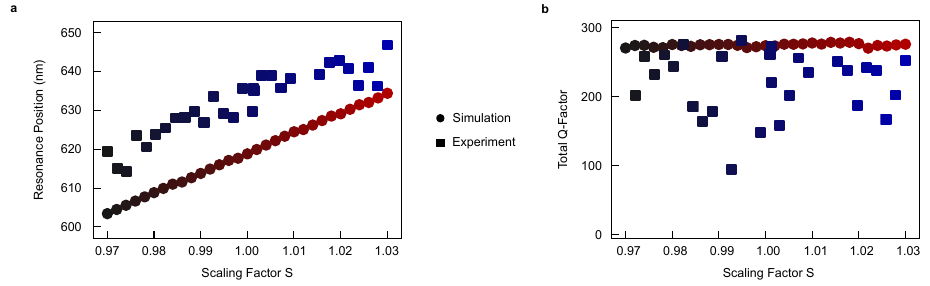}
    \caption{Introduction of Scaling Factor. 
   Comparison of \textbf{a} resonance position and \textbf{b} total Q-factor for simulated and experimental spectra for a radial qBIC with a radius of 3.9 \textmu m scaled with scaling factors ranging from 0.97 to 1.03.}
    \label{fig:Scaling}
\end{figure}

The experimental and simulated implementation of this can be seen in Fig.~\ref{fig:Scaling}, where in both cases a base ring of \textit{R}~=~3.95\textmu m radius, \textit{l}~=~920 nm resonator length, and \textit{d}~=~60 nm gaps is scaled linearly with scaling factors \textit{S} ranging from 0.97 to 1.03 in a total of 30 equidistant steps. The unscaled parameters of unit cell and resonator height are \textit{N}~=~68 and \textit{h}~=~165 nm, respectively. The simulated results match the desired continuous and linear shift of resonance position with constant $Q$-factor. 
The experimental data follows the same trend in resonance position but with a red-shifted baseline, most likely due to anisotropic etching yielding an unaccounted-for increase in resonator volume. The quality factors of the experimental radial qBICs vary between 100 and 300. Nevertheless, the implementation of such a scaling factor allows very accurate shifting of the resonance of separate radial qBICs across a desired spectral range while maintaining the intended resonance lineshape. Furthermore, a substantial number of independently nanoscale-tuned parameters are consolidated into a singular tuning parameter, allowing for more straightforward analysis and discussion.

\clearpage
\section{Supplementary Note 5 - Numerical Investigation of Strong Coupling}
\begin{figure}[h]
    \centering
    \includegraphics[width=1\linewidth]{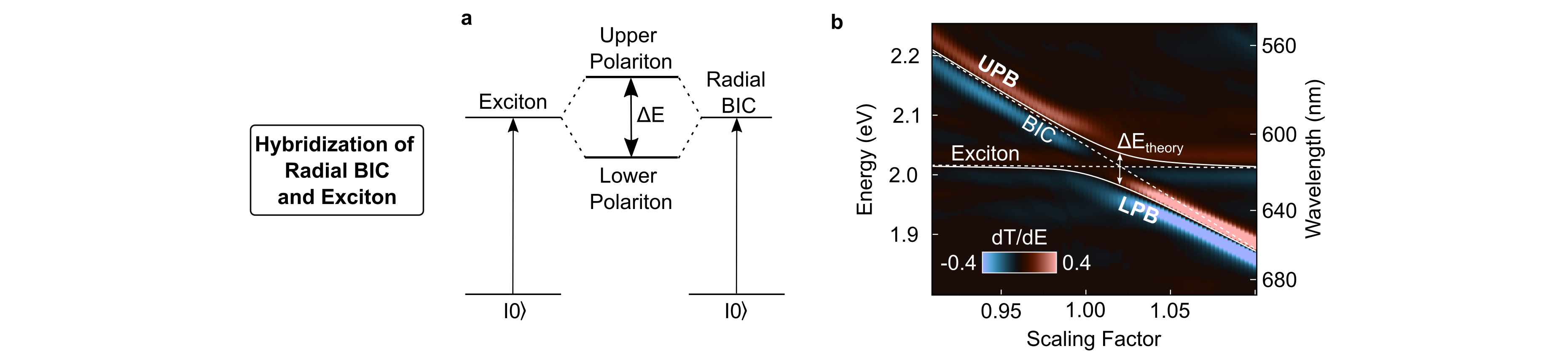}
    \caption{Strong Coupling. 
     \textbf{a} Simplified energy level diagram of self-hybridization of excitons and radial BIC into polaritons. 
     \textbf{b} Derivative of the transmittance (dT/dE) for simulated scaling increase showing anticrossing behavior.}
    \label{fig:SC}
\end{figure}

Spectral coincidence such as in main text Fig.~ 3f points to the possibility of exciton-photon hybridization in the radial qBIC heterostructure system. A simplified energy level diagram (Fig.~\ref{fig:SC}a) illustrates the conceptual formation of hybrid polariton states arising from coherent mixing between the exciton and qBIC mode. To explore this effect, we perform numerical simulations of the transmittance derivative (dT/dE) as a function of structural scaling (Fig.~\ref{fig:SC}b). The resulting spectral dispersion exhibits an anticrossing near the exciton energy, a feature consistent with the formation of hybrid light-matter states. These results demonstrate the potential of radial qBIC resonantors to generate self-hybridized exciton-polaritons, opening the door for future experimental studies of strong coupling in such radially symmetric systems.

\newpage

\section{Supplementary Note 6 - Illustration of Fabrication Workflow}
\begin{figure}[ht!]
    \centering
    \includegraphics[width=1\linewidth]{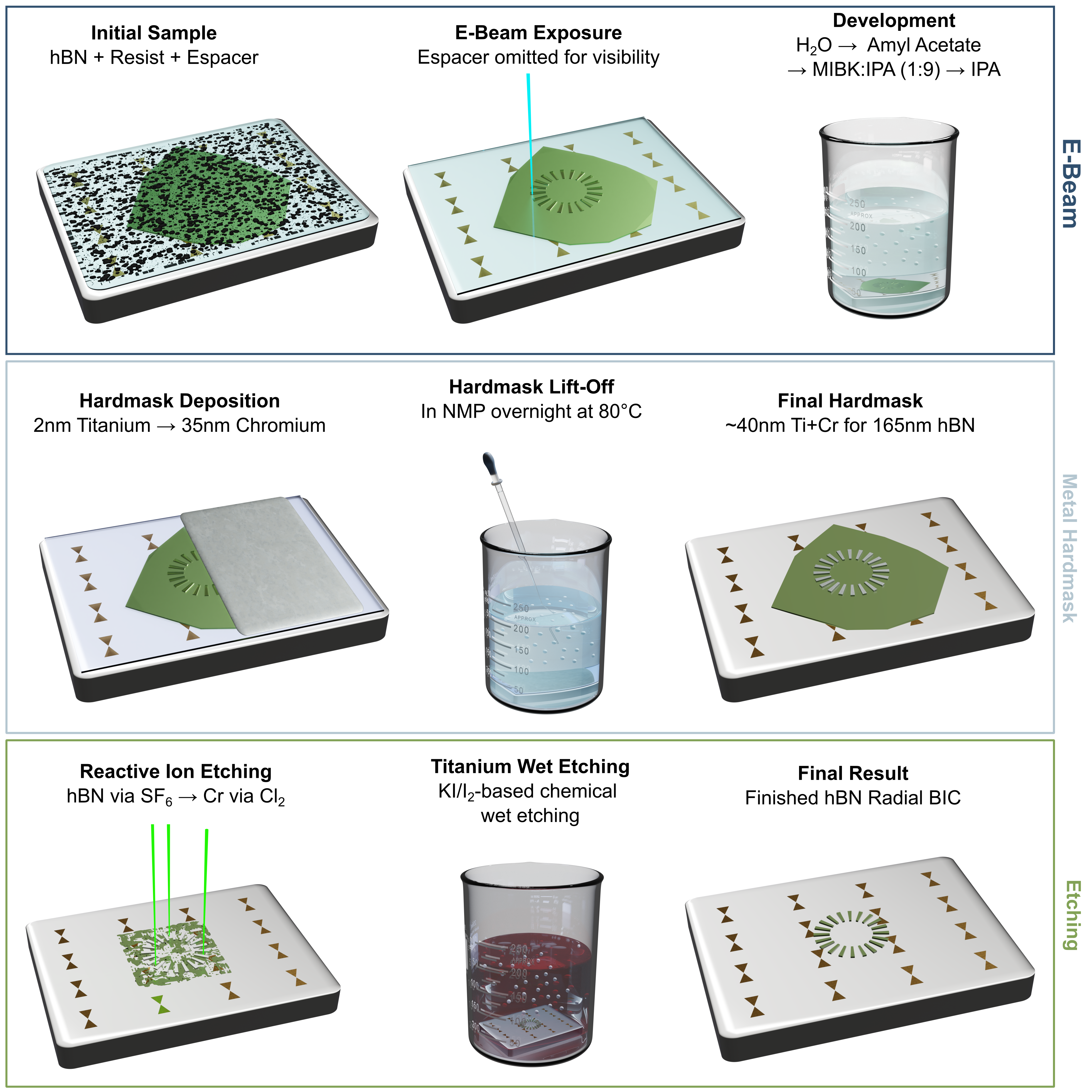}
    \caption{Fabrication Workflow. 
    Graphic outline of the workflow for e-beam exposure, subsequent hardmask deposition and etching resulting in final hBN radial qBIC platform.}
    \label{fig:Fabrication}
\end{figure}

\clearpage  
\section{Supplementary Note 7 - Sketch of Experimental Setup}

\begin{figure}[ht]
    \centering
    \includegraphics[width=0.35\linewidth]{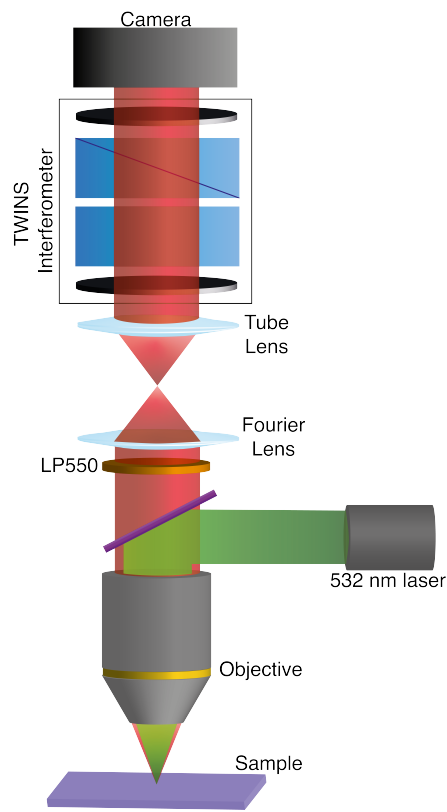}
    \caption{Sketch of the Fourier-space hyperspectral microscope. 
    A 532\,nm CW laser is used to illuminate the sample, filtered by a dichroic mirror and long-pass filter. A $100\times$ objective (NA = 0.75) collects the emission. A Fourier lens and TWINS interferometer image the back focal plane onto the camera. Details on the technique can be found in ~\cite{brida2012phase, genco2022k}.}
    \label{fig:kSpace}
\end{figure}

\end{document}